\documentstyle[12pt]{article}
\title{\bf $\omega$ Centauri: \\Nucleus of a Milky Way Dwarf Spheroidal?}
\author{S.~R.~Majewski $^1$\thanks{Visiting Associate, Carnegie Observatories; NSF CAREER and Packard Fellow, Cottrell Scholar.},
R.~J.~Patterson $^1$, D.~I.~Dinescu $^1$, \\
W.~Y.~Johnson $^1$, J.~C.~Ostheimer $^1$, W.~E.~Kunkel $^2$ and C.~Palma $^1$\\
\vspace{0.2cm}\\
\normalsize $^1$University of Virginia, Charlottesville, Virginia, USA \\
\normalsize $^2$Las Campanas Observatory, Carnegie Institution of Washington, La Serena, Chile \\}
\date{\mbox{}}

\begin{document}
\maketitle
\pagestyle{empty}
%
%
\def\bull{\vrule height .9ex width .8ex depth -.1ex}
\makeatletter
\def\ps@plain{\let\@mkboth\gobbletwo
\def\@oddhead{}\def\@oddfoot{\hfil\tiny\bull\quad
``The Galactic Halo~: from Globular Clusters to Field Stars'';
35$^{\mbox{\rm th}}$ Li\`ege\ Int.\ Astroph.\ Coll., 1999\quad\bull}%
\def\@evenhead{}\let\@evenfoot\@oddfoot}
\makeatother
%
%
\def\beginrefer{\section*{References}%
\begin{quotation}\mbox{}\par}
\def\refer#1\par{{\setlength{\parindent}{-\leftmargin}\indent#1\par}}
\def\endrefer{\end{quotation}}
\def\plotfiddle#1#2#3#4#5#6#7{\centering \leavevmode
\vbox to#2{\rule{0pt}{#2}}
\includegraphics{#1}}

\def\deg{\hbox{$^\circ$}}
%
%
{\noindent\small{\bf Abstract: We derive the metallicity distribution
    for the globular cluster $\omega$ Centauri, and combine this with
    both a new determination of its orbit as well as its other unique
    properties to argue that $\omega$ Cen may be the remains of a
    formerly larger dwarf galaxy that has undergone substantial tidal
    stripping.}
%
%
\section{Introduction}

We focus on two aspects of $\omega$ Cen that are rather uncharacteristic
of globular clusters: its metallicity distribution and its orbit. Of
course, $\omega$ Cen has other properties, such as mass and shape, that
lend support to the idea that it is a ``transitional'' object with
properties between those of dwarf galaxies and globular clusters.  A
dependence of kinematics on metallicity (Norris et al.\ 1997, ``NFMS'') and an
apparent dependence of spatial distribution on metallicity within the
cluster (Jurcsik 1998), bring even more complexity to the question of
$\omega$ Cen's origin.

Given the spread in our metallicity distribution function (MDF), as
derived from Washington + DDO photometry, as well as the orbital
parameters of $\omega$ Cen and the other unusual aspects named above, we
propose that this object is the remains (nucleus) of a once larger
satellite dwarf galaxy that has been substantially reduced by tidal
stripping.

\section{Photometric Catalogue}

Our color magnitude diagram (CMD; Fig.\ 1), containing over 130,000
stars, was constructed from CCD images taken on UT March 15-17, 1997
with the Swope 1-m at Las Campanas Observatory through the Washington
$M$ and $T_2$ filters and covering $\sim 1$ deg$^2$ around $\omega$ Cen.
We also have images in the gravity-sensitive, intermediate-band
$DDO51$ filter, which allows us to discriminate foreground field dwarfs
from giants (Majewski et al.\ 1999, ``MOKP'').  We proceed by limiting
our analysis to all stars with magnitude errors $<0.05$ mag.  Next, by
using the $(M-DDO51)_o$ color, which is sensitive to the Mg features
around 5150\AA, we weed out dwarfs from giants (MOKP; Fig.\ 2); this
exercise is particularly helpful for finding  stray, metal-rich
$\omega$ Cen giants among Galactic foreground stars, and  allows us to
explore the cluster to low densities without fear of Galactic dwarf
contamination.  
\indent Metallicities are obtained from the secondary dependence of the $DDO51$
filter on metallicity, and 

\noindent the metallicity scale is obtained (Fig.\ 2;
see MOKP) by adjusting the Paltoglou \& Bell (1994) synthetic photometry
to stars observed spectroscopically by Suntzeff \& Kraft (1996, ``SK'').

\begin{figure}
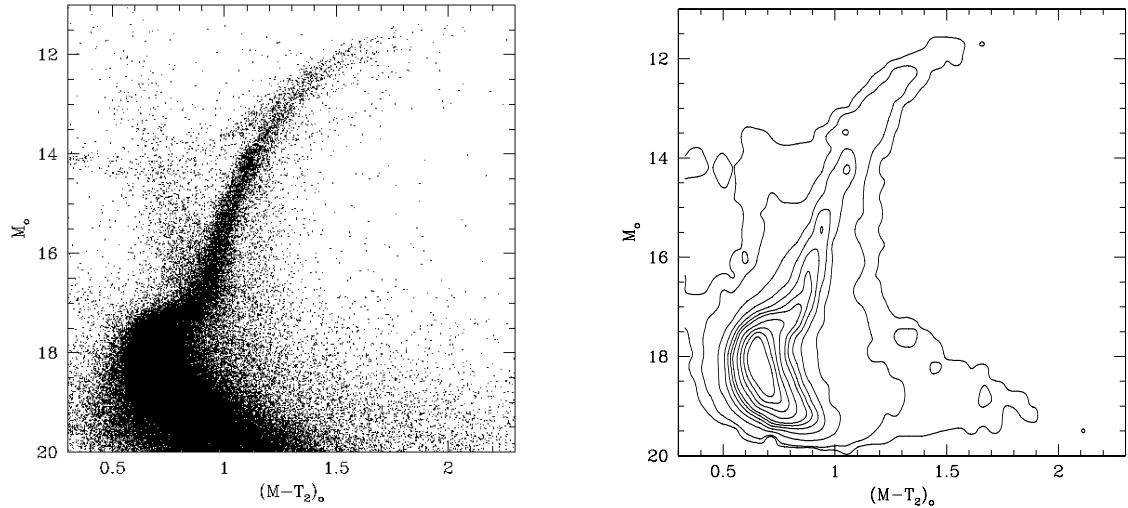

\plotfiddle{Majewski1a.eps}{30truemm}{0}{35}{35}{-230}{-100}
\plotfiddle{Majewski1b.eps}{0truemm}{0}{50}{70}{-45}{-197}
\plotfiddle{box.eps}{0truemm}{0}{35}{35}{-0}{-47}
\caption{$\omega$ Cen color-magnitude (left) and Hess diagrams (right) of the giant branch region.}
\end{figure}

The ``raw" MDF is shown in Fig.\ 3
({\it dashed + solid lines}).  As a check on the possibility of field
{\it giant} contamination in our sample, we obtained radial velocities
from Du Pont 2.5-m spectra for all of the giant stars with $1.35 <
(M-T_2)_o < 1.40$. The velocities of this sample show a rather sharp division in that all
but one of 31 giant stars with photometric [Fe/H] $> -1.19$ are {\it
  not} members of $\omega$ Cen (and the one member in this interval has
a photometric [Fe/H] = $-1.14$), while all 32 giants with lower
photometric [Fe/H] {\it are} $\omega$ Cen members.  Presumably the
non-member giants are from the field disk population.  We have applied this
near-step-function membership probability distribution to correct the
MDF, as shown in Fig.\ 3 ({\it solid lines}). Our technique becomes
insensitive outside of $-2.23<$[Fe/H]$<-0.47$, and we show outliers at
the edge of our MDF.

\section{Discussion}

We discuss several particularly germane features of $\omega$ Cen:

\underline{MASS and SHAPE} \hskip5pt The large mass (M $\sim 7$x$10^{6}$
M$_{\odot}$, the most massive Galactic globular) and flatness ($\epsilon
= 0.121$, Geyer et al.\ 1983, compared with a median of 0.006 and mode
of 0.03 for a sample of 100 globulars, White \& Shawl 1987) are the most
obvious characteristics that distinguish $\omega$ Cen from other
globulars in the Milky Way.  That the major axis of $\omega$ Cen is
almost perpendicular to the orbit derived by Dinescu et al.\ (1999) is
not a problem for our hypothesis that $\omega$ Cen may be in the process
of tidal disruption since the observed flattening is due to rotation and
probably primordial (Merritt et al.\ 1997).  Low latitude, differential
reddening problems over the scale of several tidal radii ($r_t \sim
45'$) will make a search for the likely more extended $\omega$ Cen {\it
  tidal} features difficult (see Meylan discussion of his work with Leon
and Combes in these proceedings).

\underline{ABUNDANCE} The chemical composition of $\omega$ Cen has long
been recognized as atypical for globulars.  A number of papers have
analyzed the MDF, and agree that it is at least very wide, with a tail
towards high [Fe/H] (Norris at al. 1996, SK). Our results affirm the
significant spread, with {\it a well populated MDF at all} [Fe/H] where
our technique is sensitive, but with few stars having
[Fe/H]$\mathrel{\hbox{\rlap{\hbox{\lower5pt\hbox{$\sim$}}}\hbox{$>$}}}
-1.2$.  The spread points to a complex formation process, and one that
may have left a curiously inhomogeneous spatial and kinematical
distribution of stars with different [Fe/H] (Jurcsik 1998, NFMS).

\begin{figure}
\plotfiddle{Majewski2.eps}{40truemm}{-90}{35}{35}{-240}{190}
\plotfiddle{Majewski3.eps}{0truemm}{0}{35}{35}{280}{-5}
\begin{minipage}{9cm}
\caption{Color-color diagram for separating giants and dwarfs.
  All stars from Fig.\ 1 with magnitude errors less than 0.05 mag are
  included. Lines of various [Fe/H] for giants are shown.  }
\end{minipage}
\hfill
\begin{minipage}{7cm}
\caption{Raw ({\it dashed + solid lines}) and corrected ({\it solid line}) MDF for $\omega$ Cen giants from Fig.\ 2 with $(M-T_2)_o\ge1.3$.  Stars with [Fe/H]$\le-2.23$ ([Fe/H]$\ge-0.47$) are put into edge bins.}
\end{minipage}
\end{figure}
  
\underline{AGE} Wallerstein (this conference) reports a 4 Gyr age spread
in $\omega$ Cen.  We cannot confirm this from our data: Although the
MSTO region in the Hess diagram (Fig. 1) appears to be wider than one
would expect solely from the metallicity spread observed in the giant
branch (which should have a similar width for a given single age), we are
reserved in our interpretation because of degenerate combinations of
[Fe/H] and age that can account for the MSTO spread.

\underline{ORBIT} The orbit of $\omega$ Cen is also rather unusual in
that it is relatively strongly retrograde, but of a small size, and
relatively confined to the Galactic plane.  A comparison to other
clusters is made in Fig.\ 4, which shows orbital inclination
to the Galactic plane, $\Psi$, versus the normalized
orbital angular momentum, L$_z$/L$_{z,max}$ (where L$_z$ is the actual angular
momentum, and L$_{z,max}$ the orbital angular momentum of a circular
orbit of the same total energy). The upper-left panel of Fig.\ 4 shows
all clusters with orbital data (38 clusters), with $\omega$ Cen labeled.
The other panels show different groups of clusters defined by their
horizontal branch (HB) morphology and metallicity (Zinn 1996): BHB --
blue HB clusters for their metallicity (typical first-parameter
clusters), MP-- metal poor clusters ([Fe/H]$< -1.8$), and RHB -- red HB
clusters for their metallicity (second-parameter clusters). Two
metal-rich disk clusters are also labeled in the BHB sample.

From the standpoint of the orbital trends in Fig. 4 alone, $\omega$ Cen
shows the {\it worst} fit with the very cluster group to which it
normally would be assigned, i.e., the BHB group.  A better orbital match
is had with the RHB group, however, the size (and therefore total
energy) of $\omega$ Cen's orbit is much smaller than that of the RHB
clusters (the smallest apocentric radius in the RHB group is $\sim 11$
kpc, compared to 6.4 kpc for $\omega$ Cen, Dinescu et al.\ 1999), and,
in any case, $\omega$ Cen has an unambiguous, very blue HB.  The mean
[Fe/H] of $\omega$ Cen is only slightly higher than the upper limit for
the MP group, which also might be taken to show a better match to
$\omega$ Cen's orbital properties; however, the one cluster (NGC 6779)
in the MP group with an orbit almost as extreme as $\omega$ Cen is
actually about 0.3 dex more metal poor (at [Fe/H] = $-1.94$).

A possible scenario producing a strongly retrograde orbit with small
apocentric radius and closely confined to the Galactic plane is that
$\omega$ Cen originated in, or as, a massive satellite (that happened to
have an originally retrograde orbit) that experienced significant
dynamical friction from the dark halo, and subsequently also from the
disk, {\it before} it underwent significant disruption.  The present mass of
$\omega$ Cen is unlikely to have generated strong enough dynamical
friction to modify its orbit to its presently small apocentric radius.
This proposed scenario resembles the N-body models of Walker et al.\ 
(1996, ``WMH''), which show that inclined orbits of satellites first
sink towards the Galactic plane before substantial radial decay of the
orbit, and significant disruption, ensue.  In their models, retrograde
satellites face considerable shredding as they pass through the disk.

\begin{figure}
\plotfiddle{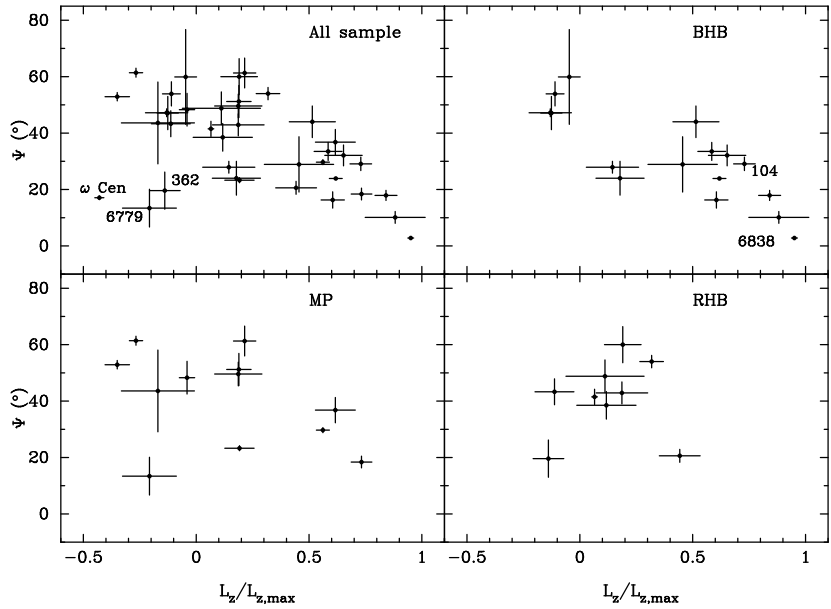}{65truemm}{-90}{45}{45}{-180}{250}
\caption{Orbital parameters for classes of globular clusters.}
\end{figure}

In light of the characteristics that make $\omega$ Cen unique among
Galactic globular clusters, we suggest that $\omega$ Cen may be the
surviving nucleus of a dwarf spheroidal that has been accreted by the
Galaxy. The abundance, age and orbital parameters of $\omega$ Cen are
consistent with such an origin, and a present-day counterpart of this
process can be seen in the Sagittarius dSph system, for which the
globular cluster M54, which has a mass similar to $\omega$ Cen, has been
purported to be the nucleus (e.g., Bassino \& Muzzio 1995).  The idea
that some globular clusters may be the remains of dwarf nucleated
ellipticals (Zinnecker et al.\ 1988, Freeman 1993) may find
representatives in M54 and $\omega$ Cen.  Moreover, the WMH models
predict that the surviving remnants of merging satellites form compact,
globular cluster-like bodies as they burrow through the disk. Under this
scenario, the more diffuse outer envelope of proto-$\omega$ Cen would
have been accreted by the Milky Way and contributed to the field halo
and thick disk populations. In addition, a disintegrating proto-$\omega$
Cen plowing through the disk backwards may have contributed substantial
disk heating, perhaps aiding thick disk formation (see WMH).

%
%
\section*{Acknowledgements}
This research was supported in part by an NSF CAREER Award, a David and Lucile
Packard Foundation Fellowship, and a Cottrell Scholar Award from the Research 
Corporation to SRM.  
%
%
 
\beginrefer

\refer Bassino, L. P. \& Muzzio, J. C. 1995, Observatory, 115, 256

\refer Dinescu D. I., Girard T. M. \& van Altena W. F., 1999, AJ 117, 1792

\refer Freeman K. C., 1993, ASP Conf. Series, Vol. 48, 608

\refer Geyer E. H., Nelles, B. \& Hopp U., 1983, A\&A 125, 359

\refer Jurcsik J., 1998, ApJ 506, L113

\refer Majewski S. R., Ostheimer J. C., Kunkel W. E., Patterson, R. J., 1999, AJ, {\it submitted} (MOKP)

\refer Merritt D., Meylan G. \& Mayor M., 1997, AJ 114, 1074

\refer Norris J. E., Freeman K. C., Mayor M. \& Seitzer P., 1997, ApJ, 487, L187 (NFMS)

\refer Norris J. E., Freeman K. C. \& Mighell K. J., 1996, ApJ 462, 241

\refer Paltoglou G. \& Bell R. A., 1994, MNRAS 268, 793

\refer Suntzeff N. B. \& Kraft R. P., 1996, AJ 111, 1993 (SK)

\refer Walker, I. R., Mihos, J. C. \& Hernquist, L. 1996, ApJ, 460, 121 (WMH)

\refer White R. E. \& Shawl S. J., 1987, ApJ 317, 246

\refer Zinnecker, H., Keable, C. J., Dunlop, J. S., Cannon, R. D. \& Griffiths, W. K.\ 1988, in IAU Symposium 126, ed. J.E. Grindlay \& A.G. Davis Philip (Dordrecht: Kluwer), p. 603.

\refer Zinn R. 1996, in {\it The Galactic Halo...Inside and Out}, ASP
Conf. Ser. Vol. 92, eds. H. Morrison, A. Sarajedini, (San Francisco: ASP) p. 211

\endrefer           
\end{document}